# Lorentz meets Fano spectral line shapes: A universal phase and its laser control

Christian Ott[1], Andreas Kaldun[1], Philipp Raith[1], Kristina Meyer[1], Martin Laux[1], Jörg Evers[1], Christoph H. Keitel[1], Chris H. Greene[3], and Thomas Pfeifer[1,2]*

[1]Max-Planck Institute for Nuclear Physics, 69117 Heidelberg, Germany
[2]Center for Quantum Dynamics, Ruprecht-Karls-Universität Heidelberg, 69120 Heidelberg, Germany
[3]Department of Physics, Purdue University, West Lafayette, Indiana 47907, USA

*Correspondence to: thomas.pfeifer@mpi-hd.mpg.de

**Abstract:**
Symmetric Lorentzian and asymmetric Fano line shapes are fundamental spectroscopic signatures that quantify the structural and dynamical properties of nuclei, atoms, molecules, and solids. This study introduces a universal temporal-phase formalism, mapping the Fano asymmetry parameter $q$ to a phase $\varphi$ of the time-dependent dipole-response function. The formalism is confirmed experimentally by laser-transforming Fano absorption lines of autoionizing helium into Lorentzian lines after attosecond-pulsed excitation. We also prove the inverse, the transformation of a naturally Lorentzian line into a Fano profile. A further application of this formalism amplifies resonantly interacting extreme-ultraviolet light by quantum-phase control. The quantum phase of excited states and its response to interactions can thus be extracted from line-shape analysis, with scientific applications in many branches of spectroscopy.

In spectroscopy, the symmetric Lorentzian line shape is one of the most ubiquitous spectral features. It corresponds to an exponentially decaying excited state with a finite life time. In the cases of detection by electromagnetic radiation, it is the temporal dipole response—the time-dependent dipole moment of the system after an infinitesimally short (Dirac delta function) excitation—of the medium that gives rise to spectral lines and their shapes observed in fluorescence or absorption. If a continuum of states is excited, this temporal dipole response corresponds also to a delta function, which is the superposition of a continuous spectrum of emitting dipoles at all frequencies.

Asymmetric Fano absorption line shapes emerge when discrete excited states are coupled to a continuum of states (*1,2*), which is a general phenomenon throughout nuclear (*3*), atomic (*4-6*), and solid-state physics (*7-10*), as well as molecular spectroscopy in chemistry (*11*). As a result of this discrete–continuum coupling mechanism, the temporal dipole response function is not just the sum of the exponentially decaying and delta-like dipole responses of the isolated state and continuum, respectively. The exponential dipole response is shifted in phase with respect to the Lorentzian response, which is the origin of the asymmetric line shape of the Fano resonance. By a mathematical transformation (see Methods (*12*) section 1) similar to the one recently conducted for a classical Fano oscillator (*13*), we map this phase shift $\varphi$ in the time domain into the *q*- parameter, which was introduced by Ugo Fano (*1,2*) and thereafter used to characterize and quantify the asymmetric Fano line shape. The cross section at photon energy $E = \hbar\omega$ is given in terms of $q$ by:

$$\sigma_{Fano}(E) = \sigma_0 \frac{(q+\varepsilon)^2}{1+\varepsilon^2}, \quad \varepsilon = \frac{E-E_0}{\hbar(\Gamma/2)}$$

Eq. (1)

in which $E_0$ and $\Gamma$ denote the position and width of the resonance, respectively and $\hbar$ denotes the reduced Planck constant, and $\sigma_0$ is the cross section far away from the resonance.

In general, the absorption cross section $\sigma_{abs}$ is proportional to the imaginary part of the index of refraction, which in turn is directly related to the polarizability (*5*) and thus to the frequency-dependent dipole response function $d(E)$,

$$\sigma_{abs}(E) \propto \text{Im}(d(E)).$$

Eq. (2)

Via the Fourier transform, $d(E)$ is connected to the time-dependent linear response $\tilde{d}(t)$ of the medium after a delta-like excitation pulse. For a Lorentzian spectral line shape of width $\Gamma$, $\tilde{d}_{Lorentz}(t)$ is an exponentially decaying function of time with time constant $\tau = 1/\Gamma$. For the Fano resonance, equating $\sigma_{abs} = \sigma_{Fano}$ (Eq. (1) and Eq. (2)) and using causality results in the following expression for the time-dependent response function (see Methods (*12*) Section 1 for the derivation):

$$\tilde{d}_{Fano}(t) \propto c\delta(t) + \exp\left(-\frac{\Gamma}{2}t + i\left(-\frac{E_0 t}{\hbar} + \varphi(q)\right)\right)$$

Eq. (3)

i.e. an exponentially-decaying dipole moment shifted in phase by

$$\varphi(q) = 2\arg(q-i) \quad \text{and in turn} \quad q(\varphi) = i\left(\frac{1+\exp(i\varphi)}{1-\exp(i\varphi)}\right) = \frac{\sin\varphi}{\cos\varphi - 1}$$

Eq. (4)

For the special case of $\varphi=c=0$, the Lorentzian dipole response function is obtained. This correspondence is illustrated in Fig. 1 and Fig. 2.

In the context of configuration interaction of electronic states, as considered in Fano's original theory, this phase shift $\varphi$ can be understood as the consequence of the strong coupling of continuum and discrete quantum states. Originally this coupling was formulated in terms of the $q$-parameter as shown in Eq. (1). The reformulation here provides a physically intuitive picture in terms of the dipole-response function, and unveils the universal nature of the phase shift $\varphi$. In particular, an important implication of this mapping between $q$ and the phase $\varphi$ is the fact that any phenomenon that shifts the dipole evolution of a system out of phase with an initial excitation, can be used to modify the $q$-parameter, e.g. transforming a Lorentzian into a Fano line-shape profile (*14*) and *vice versa*, and thus to control the absorption process in general.

Since their invention, laser light sources have evolved technologically in a revolutionary way, now providing few-cycle light fields in the visible and near-infrared (NIR), and 'delta-like' attosecond pulses (*15*) in the extreme ultraviolet (XUV) spectral regions. Having these optical precision tools at hand, here we ask the scientific questions: Can we use lasers to "quantum simulate" Fano resonances? To what extent can absorption, in general, be controlled? Answers to these questions could help to understand and to formulate new descriptions of electron–electron interaction dynamics on one hand and allow for novel applications such as x-ray or even $\gamma$-ray lasers on the other hand.

In the following, we first experimentally prove and apply the above Fano-$q$/dipole-phase correspondence to the paradigm two-electron system of helium (to which Fano's theory was originally applied in his 1961 formulation) to switch a Fano line shape into a Lorentzian line shape and *vice versa*, by introducing an additional phase with a laser field.

In the experiment, broadband attosecond-pulsed XUV light (see Methods (*12*) section 3) is sent through a sample of He atoms at a density of ~100 mbar, serving as a delta-like excitation. The transmitted spectrum (attosecond pulse + dipole response) is resolved by a flat-field grating spectrometer in the vicinity of the sp$_{2n+}$ doubly-excited state resonance series and detected with a backside-illuminated x-ray CCD camera. In addition, a collinearly propagating 7-fs NIR laser pulse of controlled intensity is passing through the sample at a fixed delay of ~5 fs following the attosecond pulse. The intensity of the NIR pulse was controlled by opening and closing an iris in the beam path. When the NIR pulse is absent, the commonly-known Fano line shapes are observed in the absorption spectrum (Fig. 3A). With a laser field present, we observe strong modifications of the spectral features (*16*). Most strikingly, at a laser intensity of $(2.0\pm1.0)\times10^{12}$ W/cm$^2$, the asymmetric Fano profiles are replaced by symmetric Lorentzian line shapes in the spectrum (Fig. 3B). Assuming a ponderomotive shift of these states in the laser field during the pulse duration (see Methods (*12*) section 2), we calculate a total phase shift of $(-0.35\pm0.18)\pi=(-1.1\pm0.5)$ rad imprinted by the NIR pulse right after the excitation. From Eq. (4), the dipole phase $\varphi$ of the original Fano resonance sp$_{24+}$, exhibiting a $q$ of $-2.55$ (Ref. (*6*)) amounts to $\varphi = 0.24\pi$ (0.75 rad). The two phases add up and the NIR pulse thus compensates the original (Fano) dipole phase of the state and shifts it back to near zero $(-0.11\pm0.16)\pi$ (Fig. 2), producing the Lorentzian absorption line (Fig. 3B). It is important to note that the life time of the considered states is on the order of 100 fs or longer and not affected by the much shorter NIR pulse. In this limit, the phase modification right after excitation can be considered "kick-like" (impulsive).

To provide additional evidence for the validity of this concept, we experimentally apply the time-domain Fano-phase framework to singly-excited states in He. For these states the decay proceeds fully radiatively with no interfering continuum, and thus normally results in Lorentzian absorption line shapes (*17*). Here, we show the inverse transformation to the case above: the transformation of an initially Lorentzian atomic absorption line into a Fano resonance.

We use the same experimental configuration as above, except with the spectrometer set to resolve lower photon energies in the vicinity of 24 eV. When the laser field is absent, we record the well-known absorption spectrum of helium in the 1s$np$ series just below its first ionization threshold at 24.6 eV. When the laser intensity is set to $(2.1\pm1.1)\times10^{12}$ W/cm$^2$, the symmetric Lorentzian line shapes are replaced by asymmetric Fano line shapes. Again calculating the ponderomotive phase shift induced by the laser at this intensity results in $\varphi=(-0.38\pm0.19)\pi=(-1.2\pm0.6)$ rad. According to the time-domain phase formalism (Eq. (4)), this implies a Fano $q$ of +1.49 with experimental error as indicated by the red-shaded area in Fig. 2, which is in agreement with the measured line shape (Fig. 3D).

In addition to the change of the line profile, here we can also control the sign of the absorption by varying only the phase $\varphi$. Thus gain can be optically induced solely by modifying the phase of the polarization response after perturbative excitation in the absence of additional amplitude control or even population inversion of the excited state. This becomes clear by starting out from the dipole response of a Lorentzian absorption line (with an unperturbed $\varphi=0$), where the laser-controlled tunable phase $\varphi$ is again introduced to arrive at

$$\tilde{d}(t) \propto \exp\left(-\frac{\Gamma}{2}t + i\left(-\frac{E_0 t}{\hbar} + \varphi\right)\right)$$

Eq. (5)

resulting in

$$\sigma_{abs}(E) \propto \mathrm{Im}(d(E)) = \mathrm{Im}\left(-\frac{\varepsilon}{1+\varepsilon^2}e^{i\varphi} + i\frac{1}{1+\varepsilon^2}e^{i\varphi}\right) \text{ with } \varepsilon = \frac{E-E_0}{\hbar(\Gamma/2)},$$

Eq. (6)

which now can become negative when $\varphi$ departs from its original (no laser) value of 0. This is because the negatively-valued (for $\varepsilon<0$) dispersive term $\frac{\varepsilon}{1+\varepsilon^2}$ rotates, in the complex plane, by the angle $\varphi$ and acquires an imaginary part. The phase $\varphi$ can thus be interpreted as a laser-controllable mixing angle between dispersion (the real part of $d(E)$) and absorption.

To confirm this control and transformation of absorption into gain, we acquired a spectrum spatially resolved across the vertical axis, with and without laser control (Fig. 4). In the presence of the laser field, it clearly shows an enhancement of EUV light intensity at the resonance positions of the singly-excited states of helium which are normally absorbing (in the absence of the laser field). The same mechanism can, in the future, be readily applied to hard-x-ray or even γ-ray transitions with much longer lifetimes, where the $\varphi=\pi$ phase flip could instead be achieved by other means, such as nanosecond-pulsed magnetic fields (Zeeman shift) or a physical displacement of a solid-state absorber ([18]).

Having this experimental confirmation, Eq. (6) thus allows to generally interpret effects such as electromagnetically-induced transparency (EIT) ([19,20]) ($\sigma_{abs}(E) \propto \text{Im}(d(E)) = 0$) and 'lasing without inversion' ([21,22]) ($\sigma_{abs}(E) < 0$) in a unified intuitive picture. It also connects to general dispersion control ([14,22]) with short-pulsed fields. The formalism developed here is universal and may help to provide intuitive physical pictures also for cases of non-resonant amplification of light recently experimentally observed in helium atoms ([23]) and the laser coupling of discrete autoionizing states ([24,25]). For that latter situation, modifications of Fano profiles in simulated transient-absorption spectra for delta-like excitation and coupling have recently been observed and analytically described ([25]).

The existence of a direct correspondence between Fano's *q* parameter and the dipole phase of an excited state has thus been proved. As the latter is susceptible to laser fields, Fano absorption profiles can be induced in the absence of effects such as autoionization. This is important because the change of the absorption profile is in turn a measure of the induced phase shift of a complex quantum-mechanical state amplitude in a laser field, with numerous applications in spectroscopy and quantum-state holography. The now well-understood phase-to-*q*-correspondence allows to map out the coupling of these states to laser fields or other interactions, providing information especially when the coupling mechanisms are more complex than just a ponderomotive coupling which is considered here to introduce the principle. These additional couplings are already expected in helium atoms for the more closely-bound quantum states, or for states in which both electrons are excited to the same or similar quantum numbers, for which electron–electron interaction effects play a major role while interacting with the laser. Our additional experiments (not shown here), confirm that this Fano-phase-control mechanism is not restricted to helium, but rather exhibits its signature in other more complicated (many-electron) atoms as well. There is no reason why the mechanism should not be applicable to molecules, excitons in condensed phase or mesoscopic materials.

In conclusion, we have introduced a general mathematical concept that relates spectral absorption lines to a universal and physically meaningful, measurable and controllable temporal phase $\varphi$ of the dipole response, which serves two important purposes: First, this phase $\varphi$ can be readily mapped to the already familiar $q$ parameter that characterizes Fano line shapes and thus provides a physically intuitive interpretation of the Fano process in the time domain. And second, this phase $\varphi$ can be directly modified by a laser field (or any other interaction, e.g. configuration interaction in Fano's theory), and thus can change absorption line shapes from symmetric (Lorentzian) to asymmetric (Fano). We have also applied the mechanism to induce gain near singly-excited helium resonances that are normally absorbing.

This formalism can, in the future, be used to fully control absorption and dispersion, including their signs, allowing the amplification of near-resonant radiation in any region of the electromagnetic spectrum. It provides time-domain access and an intuitive physical picture for understanding and controlling processes such as electromagnetically-induced transparency (EIT), lasing without inversion, and slow and fast light, with the potential to enable quantum-optics applications all the way up to the x-ray region at free-electron lasers (FELs). The notion of this universal phase will also provide routes to the understanding and controlling of coupled electronic processes across many different time scales and many scientific disciplines, involving simple atomic to complex molecular, condensed-phase, and solid-state systems. It also provides an intuitive link between quantum (configuration interaction, energy shifts of quantum states) and classical phenomena (classical light fields, phase shifts of oscillators), which could possibly spawn novel quantum-classical hybrid pictures of multi-electron dynamics.

We acknowledge helpful discussions with Robert Moshammer and Joachim Ullrich. Financial support from the Max-Planck Research Group program and the DFG (grant No. PF 790/1-1) is gratefully acknowledged. The work of CHG is supported in part by the U.S. Department of Energy, Office of Science.

Reference List

1. U. Fano, *Nuovo Cimento* **12,** 154-161 (1935).

2. U. Fano, *Phys.Rev.* **124,** 1866-1878 (1961).

3. J. M. Blatt and V. F. Weisskopf, Theoretical Nuclear Physics (Wiley, 1952).

4. R. P. Madden, K. Codling, *Phys.Rev.Lett.* **10,** 516 (1963).

5. U. Fano, J. W. Cooper, *Rev.Mod.Phys.* **40,** 441-507 (1968).

6. J. M. Rost, K. Schulz, M. Domke, G. Kaindl, *J.Phys.B* **30,** 4663-4694 (1997).

7. J. A. Fan et al., *Science* **328,** 1135-1138 (2010).

8. A. Schmidt et al., *Nature* **465,** 570-576 (2010).

9. M. Kroner et al., *Nature* **451,** 311-314 (2008).

10. A. E. Miroshnichenko, S. Flach, Y. S. Kivshar, *Rev.Mod.Phys.* **82,** 2257-2298 (2010).

11. S. H. Linn, W. B. Tzeng, J. M. Brom, C. Y. Ng, *J.Chem.Phys.* **78,** 50-61 (1983).

12. *Methods are available online*.

13. D. Riffe, *Phys.Rev.B* **84,** 064308 (2011).

14. C. Szymanowski, C. H. Keitel, B. J. Dalton, P. L. Knight, *J.Mod.Opt.* **42,** 985-1003 (1995).

15. P. B. Corkum, F. Krausz, *Nat.Physics* **3,** 381-387 (2007).

16. C. Ott et al., *arXiv:* **1205.0519v1,** [physics.atom-ph] (2012).

17. K. Ito, K. Yoshino, Y. Morioka, T. Namioka, *Physica Scripta* **36,** 88-92 (1987).

18. R. Shakhmuratov, F. Vagizov, O. Kocharovskaya, *Phys.Rev.A* **84,** 043820 (2011).

19. S. E. Harris, *Physics Today* **50,** 36-42 (1997).

20. A. Safavi-Naeini et al., *Nature* **472,** 69-73 (2011).

21. O. Kocharovskaya, *Phys.Rep.-Rev.Sect.Phys.Lett.* **219,** 175-190 (1992).

22. M. O. Scully, *Phys.Rep.-Rev.Sect.Phys.Lett.* **219,** 191-201 (1992).

23. J. Herrmann et al., *arXiv:* **1206.6208v1,** [physics.atom-ph] (2012).

24. P. Lambropoulos, P. Zoller, *Phys.Rev.A* **24,** 379-397 (1981).

25. W.-C. Chu, C. D. Lin, *arXiv:* **1211.6172v1,** [physics.atom-ph] (2012).

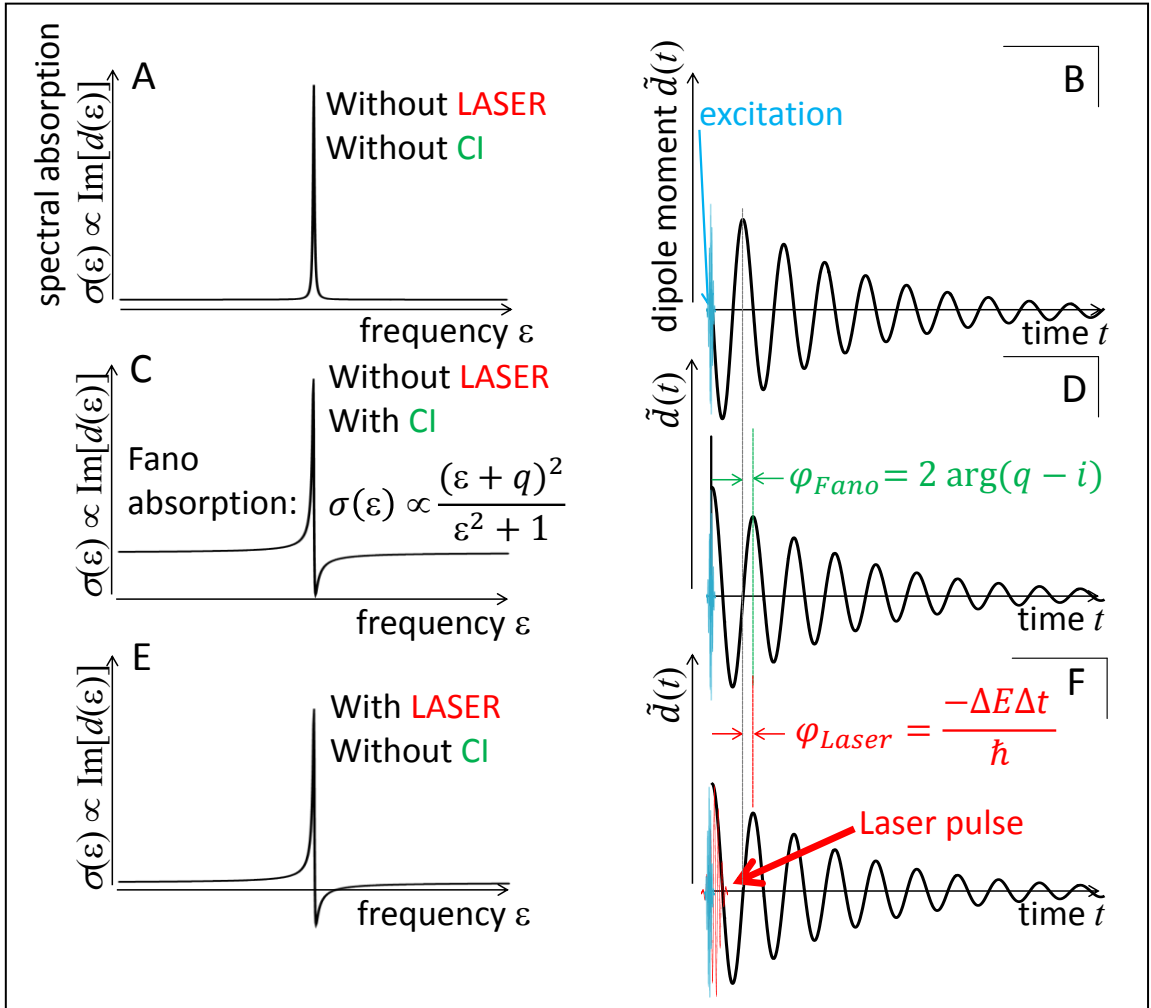

Fig. 1: **Lorentz and Fano line shapes, their temporal dipole response functions, and a universal phase that governs configuration interaction and pulsed ("kick-like") laser interaction with a state.** A: Lorentzian spectral absorption line shape, B: Lorentzian temporal dipole response function, the well-known exponential decay, C: Fano spectral absorption line shape, D: Fano temporal dipole response function, consisting of an instantaneous delta-function (continuum absorption) and an exponentially decaying response, governed by autoionization. The asymmetric line shape is understood to be caused by the phase shift $\varphi_{Fano}$ of the long-lived dipole response as compared to the Lorentzian case. This phase shift is created by configuration interaction (CI), associated with electron–electron interaction in the case of a traditional Fano resonance. E,F: Alternatively, a Lorentzian decaying state can be laser dressed impulsively ("kick-like") immediately after its excitation, inducing a phase shift $\varphi_{Laser}$ governed by the product of duration of the laser pulse $\Delta t$ and energy shift $\Delta E$. The state will then decay with a phase shift of its dipole moment (F), inducing the same Fano-like absorption line shape (E). This universal time-domain Fano-phase mechanism may hold the key to understanding many ultrafast electronically and laser-coupled processes.

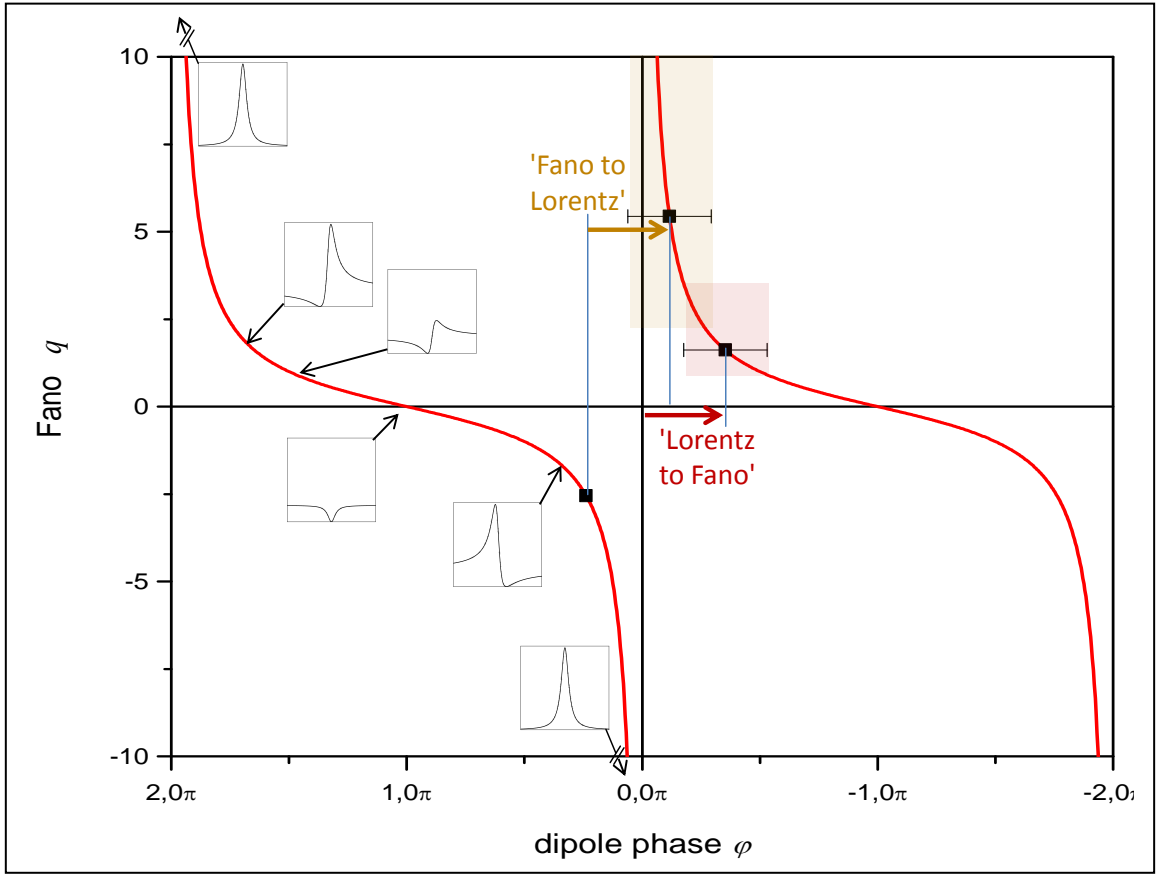

Fig. 2: **Mapping of Fano's *q* (line-shape asymmetry) parameter to the temporal response-function phase $\varphi$.** A bijective map between the two parameters is obtained in a range from [-$\pi,\pi$[, while the function is periodic in $2\pi$. Lorentzian line shapes are obtained for the extreme cases of $\varphi \to 2n\pi$ (integer *n*), corresponding to $q \to -\infty$ and $q \to +\infty$, respectively, while between these regimes Fano line shapes are obtained, with the special case of a window resonance at $\varphi=(2n+1)\pi$, $q=0$. The insets show the absorption line shapes $\sigma(\varepsilon)$ (as discussed in Fig. 1) for selected values of $q(\varphi)$. The laser interaction creates an additional phase shift (horizontal arrows) which changes the character of the observed resonance line shape. The dots represent the situations measured in the experiment and shown in Fig. 3.

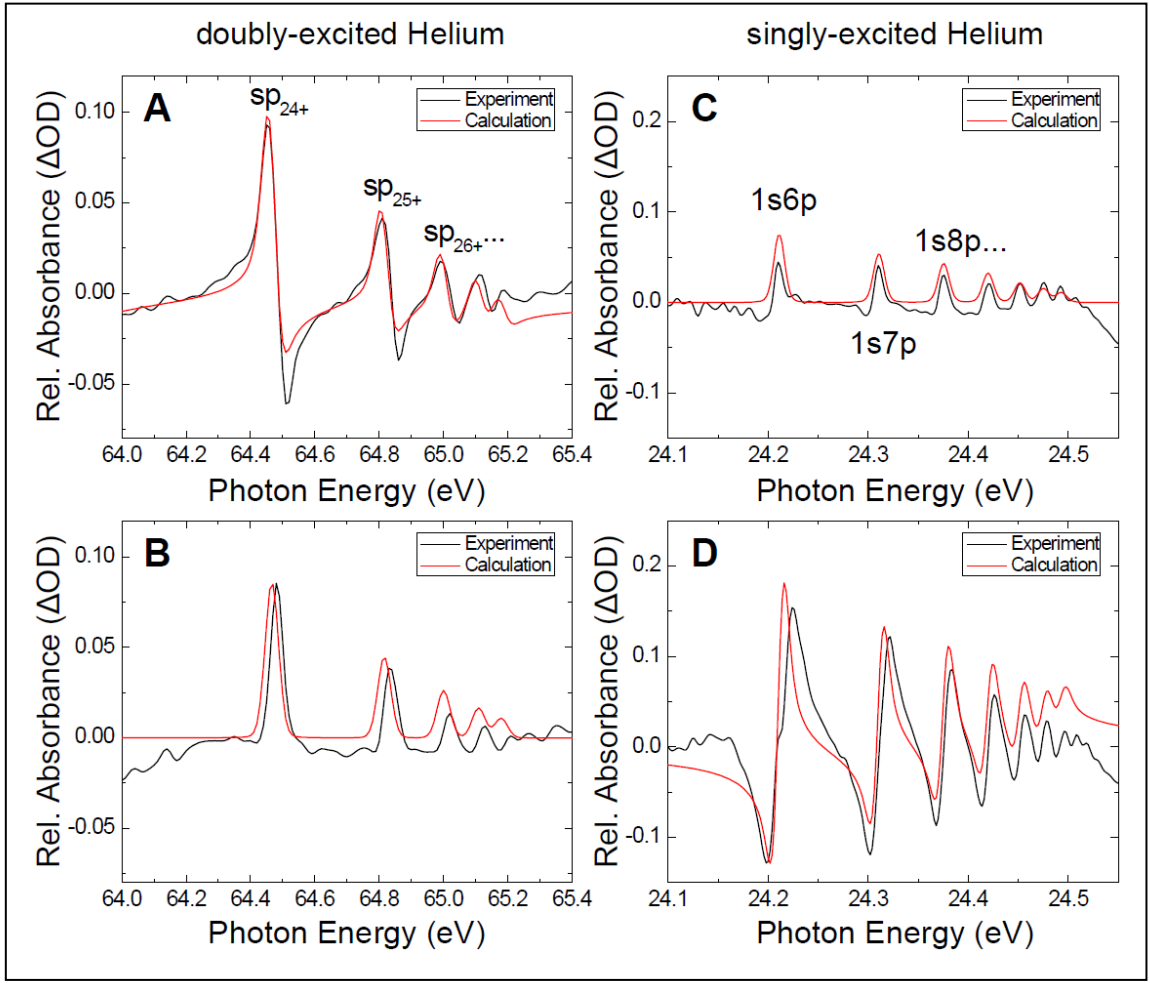

Fig. 3: **Transforming asymmetric Fano spectral absorption lines into symmetric Lorentzian absorption peaks in doubly-excited He and *vice versa* from Lorentz to Fano in singly-excited He.** A: Field-free (static) absorption spectrum of doubly-excited states of the N=2 series in He. The well-known Fano absorption profiles are observed in the transmitted spectrum of a broad-band attosecond pulse. B: When a 7-fs laser pulse immediately follows the attosecond pulsed (delta-like) excitation (time delayed by ~ 5 fs) at an intensity of $2.0\times10^{12}$ W/cm$^2$, the Fano absorption profiles are converted to Lorentzian profiles. C: Field-free (static) absorption spectrum of singly-excited He states below the first ionization threshold (24.6 eV), thus Lorentzian line shapes are visible in the attosecond pulse absorption spectrum. D: Absorption spectrum of the states in C, when the attosecond pulse is again followed by the 7-fs laser pulse, at an intensity of $2.1\times10^{12}$ W/cm$^2$. The initially Lorentzian absorption profile has been laser transformed into an asymmetric Fano profile. The solid black lines are the measurement results while the red lines are generated using tabulated values in A from Ref. (*6*) and C from (*26*), while the red line in B represents Lorentzians at the resonance positions of the original Fano lines. The red line in D shows Fano profiles with expected laser-induced *q*=1.49 (see Fig. 2) at the resonance positions of the original Lorentzian resonances. For details of the calculated profiles and the experimental data, see Methods (*12*) section 4 and 5, respectively.

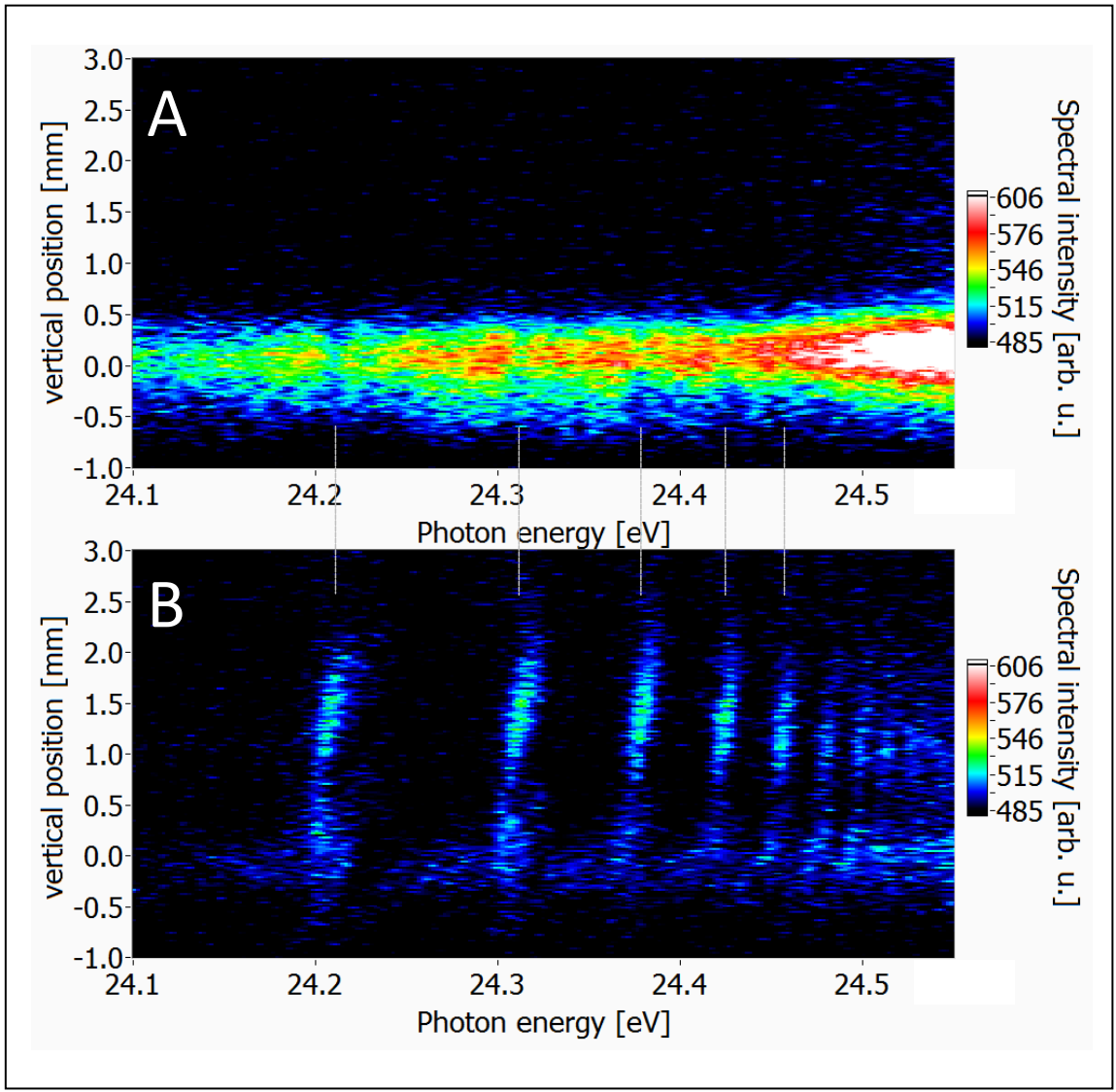

Fig. 4: **Laser-controlled amplification of resonant light in the extended ultraviolet (EUV).** A: Spectrum of transmitted EUV light without control laser, the helium resonant absorption lines can be observed as local minima in an otherwise smoothly varying spectrum centered at a vertical position of 0 mm. B: Spectrum of transmitted and amplified EUV light in the presence of the control laser. Amplification can be observed exactly at the He resonance positions corresponding to absorption in A. It is also observed slightly off-axis, which is likely due to a non-perfect angular alignment of the optical control laser with respect to the EUV beam.